# Supplementation of deep neural networks with simplified physics-based features to increase model prediction accuracy

Nicholus R. Clinkinbeard, Prof. Nicole N. Hashemi [*]

Nicholus. R. Clinkinbeard, Prof. N. N. Hashemi
Department of Mechanical Engineering
Iowa State University
Ames, IA, 50011, USA

Prof. N. N. Hashemi
Department of Mechanical Engineering
Stanford University
Stanford, CA, 94305, USA

[*] E-mail: nastaran@iastate.edu

Keywords: machine learning, deep neural networks, simplified physics, scarce data, generalizability, physics-informed

To improve predictive models for science and engineering applications, supplemental physics-based features computed from input parameters are introduced into single and multiple layers of a deep neural network (DNN). While many studies focus on informing DNNs with physics through differential equations or numerical simulation, much may be gained through integration of simplified relationships, such as available in engineering textbooks. To evaluate this hypothesis, a number of thin rectangular plates simply-supported on all edges are simulated for five materials having similar geometric parameters. With plate dimensions and material properties as input features and fundamental natural frequency as the sole output, predictive performance of a purely data-driven DNN-based model is compared with models using additional inputs computed from simplified physical relationships among baseline parameters, namely plate weight, modulus of rigidity, and shear modulus. To better understand the benefit to model accuracy, these additional features are injected into various single and multiple DNN layers, and trained with four different dataset sizes. When these physics-enhanced models are evaluated against independent data of the same materials and similar dimensions to the training sets, supplementation with simplified physics-based parameters provides little reduction in prediction error over the baseline for models trained with dataset sizes of 60 and greater, although small improvement from 19.3% to 16.1% occurs when trained with a sparse size of 30. Conversely, notable accuracy gains occur when the independent test data is of material and dimensions not conforming to the training set. Specifically, when physics-enhanced data is injected into multiple DNN layers, reductions in error from 33.2% to 19.6%, 34.9% to 19.9%, 35.8% to 22.4%, and 43.0% to 28.4% are achieved for training dataset sizes of 261, 117, 60, and 30, respectively, demonstrating attainment of a degree of generalizability.

1.   **Introduction**

Owing to its ability—in the presence of sufficient amounts of data—to predict new outcomes by determining complex relationships among various features among dataset parameters (Bianco et al., 2019), machine learning (ML) enjoys increasing application for simulating physical phenomena in science and engineering (Hashemi & Clark, 2007); (Shirsavar et al., 2021), in addition to its use as a data mining and human behavioral study tool (Shokrollahi et al., 2021). Unfortunately, models driven exclusively by data may not be conducive to generalization, particularly when such data use to generate them are limited (Anuj et al., 2014); (Nguyen et al., 2015); (Xi & Zhao, 2018). Accordingly, the presence of sparse data can lead to large prediction errors if presented with input parameters outside the range of those used for model development (Willard et al., 2020). Also limiting are the potential inaccuracies that result from uncertainties inherent to empirical measurements. Finally, purely data-driven ML models—while effective in many cases—can act as a type of "black box" where the inputs and outputs are known, but the underlying process for deriving the relationships among these remains a mystery (Adadi & Berrada, 2018). This is especially an issue when a fundamental understanding of the principles that cause certain features to drive others is crucial.

Alternatively, yet toward a similar goal of modeling complex phenomena, popular analytical and numerical approaches—such as the finite element method (FEM) and computational fluid dynamics (CFD)—are leveraged significantly in science and engineering to make behavioral predictions. Unlike with machine learning, these techniques work by solving the governing partial differential equations often rooted in first principles. For example, the FEM is a powerful tool for estimating normal modes of a structure or predicting deflections and strains during loading (Chandrupatla & Belegundu, 2002). Likewise, CFD enables the study of fluid flow and heat transfer in systems of interest through numerical solution of the Navier-Stokes equations (Pletcher et al., 1997). Notably, however, while ML suffers from lack of generalization and conceptual insight into physical processes, a conspicuous but often overlooked deficiency of analytical/numerical approaches is their inability to overcome the presence of unknown physics. In other words, since these numerical modeling methods invariably include simplifying assumptions, nontrivial phenomena driving input-output relationships in a real system are often unaccounted for in the analytical model (Blakseth et al., 2022). Additionally, physics-based numerical models may require significant time and computational resources to conduct a single simulation. When studying an array of design options, lengthy solve times may preclude adequate use of FEM, CFD, or other numerical methodologies.

Fortunately, deficiency for the physics-based model is where ML has an advantage; data derived from experimentation will—by its very nature (and within measurement limitations and uncertainties)—reflect the true physical state of a system (Blakseth et al., 2022). Therefore, benefit may be gained by successful integration of ML with physical modeling. Based on this, partnering ML techniques with numerical simulation could potentially lead to improved designs due to more efficient computation of possible solutions.

Nevertheless, available time and resources to develop a working predictive model may still be less than needed to for integration of ML with numerical simulation. Consequently, the incorporation of *simplified* physical relations with ML becomes an attractive alternative (Pawar et al., 2021). And while still a rich field of research, integration of physics into ML models has been investigated using a variety of approaches, both simple and complex. It is the goal of this study better understand the former to make more timely decisions based on available data and domain knowledge. With that, the following paragraphs discuss a few of the methodologies currently under development for pairing simplified physics and machine learning.

*Tailored loss function.* Daw et al. (2017) demonstrated the generation of tailored loss functions that include terms fueled—at least in part—by expected system physics. The additional physics-based term in the optimization process results in an increased penalty when the physical law chosen for implementation is violated. Such an approach helps to better bound the overall problem to known behavior from first principles and prior observations while still leveraging machine learning techniques.

*Reduced-order models.* Recent studies using reduced-order models (ROM) have also shown improved results over purely data-driven approaches. However, this requires effort to develop a complex model and reduce it to something manageable, which may need repeated for any changes to the system due to the unlikelihood of a single ROM having universal application (Ahmed et al., 2021). As a result, for more efficient modeling of systems, something more basic and general in nature may be of use.

*Simplified theories.* This leads to the level of physics integration investigated herein, namely, simplified theories. This methodology—presented by Pawar et al. (2021)—takes advantage of fundamental physical relationships among features to form the basis of parameter augmentation. In that study, the authors increased neural network-based model accuracy for predicting flow around an airfoil by considering fundamental fluid dynamic quantities—such as Reynolds number and angle of attack—and calculating these characteristics from existing features. Once computed, the physics-based (as opposed to purely measured) datapoints were inserted into an intermediate hidden neural network layer to supplement the original data. Additional lift and drag coefficient terms were computed based on the Hess-Smith panel method to further ground the NN-based model in physics. Their approach ultimately showed increased prediction accuracy within limitations of the underlying theory.

While perhaps not as accurate as may be found through integration with numerical-based or reduced-order models, a DNN implementation strategy that combines simplified physics with machine learning has the potential to provide solutions that are better than purely data-driven techniques, and—once a model has been built—results may potentially compute faster than for physics-only approaches.

In order to advance the use of simplified physics within a machine learning framework, the study detailed herein shows that not only can supplementing datasets with additional features generated from simple physical relations improve DNN-based model predictions (as previously demonstrated by Pawar et al.), but such physics-based information has varying impact to model predictive accuracy based on size of training dataset. Furthermore, this investigation shows that improvements are to be gained by reintegrating the same physical features into multiple DNN layers rather than simply using them once, and that different combinations of augmented layers produce notably different results, with multiple layers often outperforming those generated with physics introduced only single. Finally, infusion of a DNN with physics-based information is shown to have the greatest impact on model accuracy when used to predict behavior for data not within the scope of the original training sets. The end result is a methodology that shows at least modest improvement to DNN model accuracy and generalizability, and one that may be improved by further study and hyperparameter tuning.

## 2. Methods

The objective of this study is to examine the effects on model predictive accuracy when integrating simplified physics into a deep neural network (DNN) via data augmentation. Four specific aspects of this are investigated, as follows.

*(1) Type and number of physics relations added.* Three different simplified physics relations are used for data augmentation, both individually and in coordination. The goal is to observe whether certain individual terms or combinations thereof provide more improvement than other terms/combinations to DNN-based model accuracy.

*(2) Effects of including physics at different and/or multiple layers of the DNN.* Different cases are developed to examine how inclusion of physics affects model development when injected into each layer singularly or into multiple layers for a given DNN. The purpose of this aspect of the study is to understand if focusing on a specific DNN layer (or layers) for injecting physics has a greater influence on predictive power of the resulting model.

*(3) Effect of training dataset size on model accuracy improvement.* The third facet is of this study is to examine the impact of augmentation with simplified physics for increasing model accuracy when only small training datasets are available. Understanding that decreased training dataset size results in a potentially less-accurate model, the question becomes this: can models generated with scarce data enjoy improved accuracy when supplemented with physics?

*(4) Effect of simplified physics on predictive accuracy for inputs outside the original domain.* This final aspect is geared toward examining whether or not augmenting a DNN with simplified physics data can improve predictions when new input data is largely outside of the domain of the original. In other words, does the enhancement procedure herein create a model that can generalize better than the original? To understand the answer to this question, two basic datasets are evaluated for every combination of (1), (2), and (3) previously discussed—one having parameters within the original training dataset and the other having parameters that lie outside.

The physical phenomenon chosen for this investigation is resonance of a thin rectangular plate with four simply-supported edges, shown in Fig. 1(a). Using a DNN framework, a predictive model is developed for processing previously-measured data to develop a model that accurately predicts the fundamental resonant frequency of a new plate design using three easily measurable dimensional parameters (length, width, and thickness) and three fundamental material properties (density, Young's modulus, and Poisson's ratio). Fig. 1(b) provides a schematic of the mode shape for such a simply-supported thin rectangular plate during excitation of its fundamental resonance.

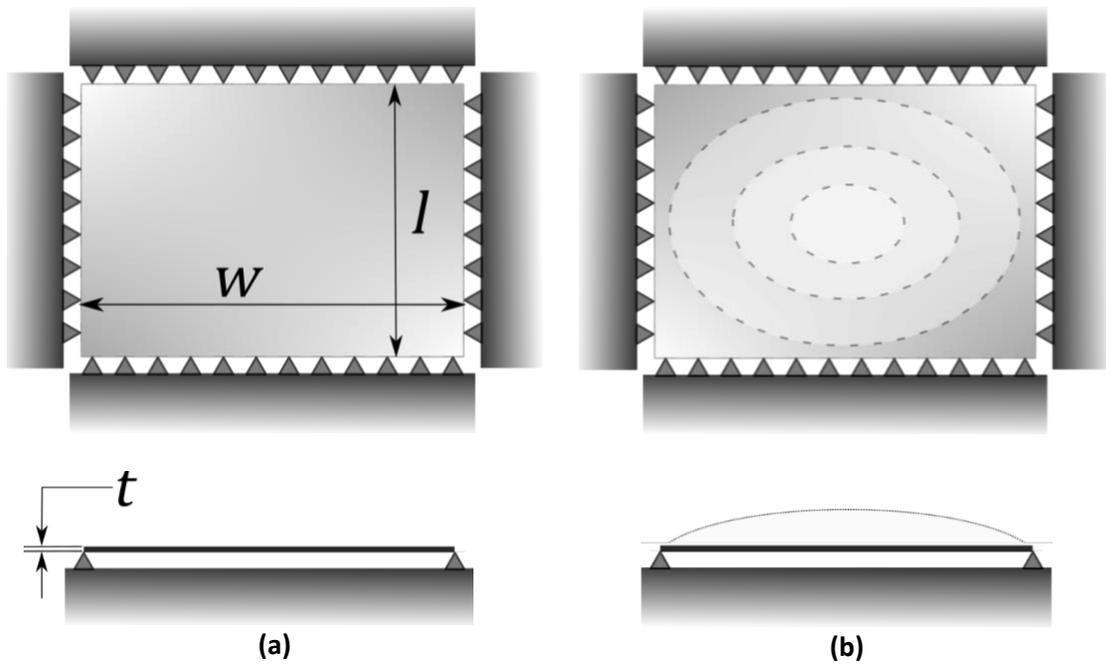

**Fig. 1.** Schematic of thin rectangular plate with simply-supported boundary conditions along all four edges, where (a) shows orthogonal (top) and planar (bottom) views of the plate having dimensions length, $l$, width, $w$, and thickness, $t$, and (b) demonstrates the mode shape for the simply-supported thin plate during excitation of its fundamental resonance.

Once predictive performance of the baseline DNN is established for four different sizes of training datasets, supplementary datapoints are generated based on the six input features (i.e., three dimensions and three material properties), and injected into the same DNN code, albeit at different layers. To assess performance of any one combination of augmenting variables/layer(s) of insertion, each DNN-based model is used to predict natural frequency of various plates containing various combinations of input features for which the models were not trained. The final performance evaluation is based on the mean error between the predicted and actual natural frequencies. Each model iteration is assessed with respect to additional input data using similar materials and dimensions to the original, as well as for a material type and dimensions not part of the original DNN training parameters.

The following paragraphs briefly describe the generation and nature of the baseline data, theory behind the development of physics-based data, and the DNN implemented to study the effects of including physics-based information.

## 2.1. Training and dataset generation

The training dataset for this study was generated through normal modes analysis using the finite element software ANSYS Version 2021, Release 1 (ANSYS, 2021). Thin rectangular plates with four simply-supported edges and predetermined—but varying—dimensions and material types were modeled, and their fundamental natural frequencies calculated. Fig. 2 shows an example finite element mesh for the simply-supported plates under study, as well as a sample contour plot of the fundamental mode shape corresponding to the frequency of interest.

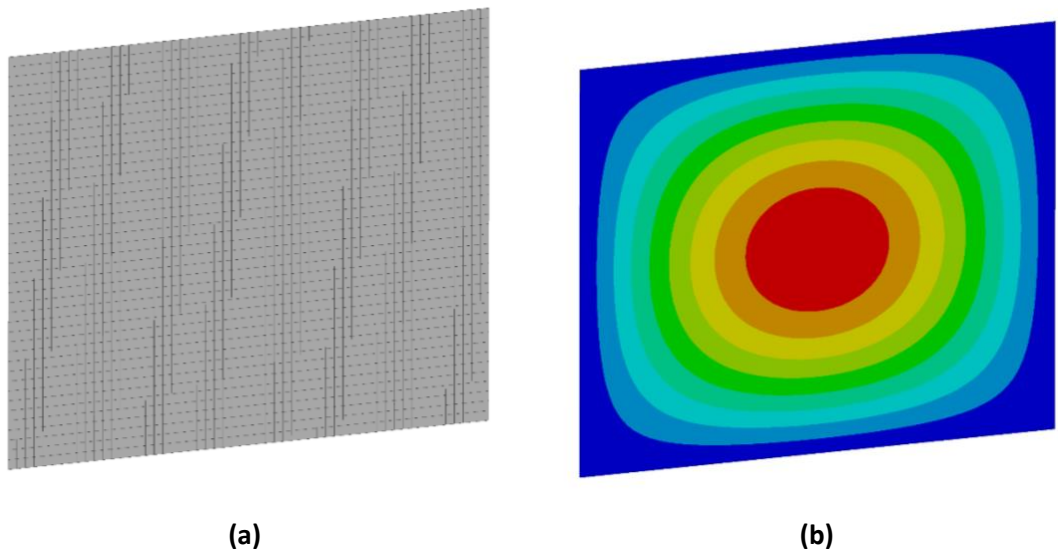

(a)          (b)

**Fig. 2.** Finite element representation of a rectangular plate with simply-supported boundary conditions along all four edges. The view in (a) shows the finite element mesh, which was developed to be relatively fine for accurate prediction of natural frequency. The view in (b) is fringe plot of the fundamental mode shape, which is similar for all combinations of dimensions and materials discussed herein.

As well as serving as inputs to the finite element models, the minimal dataset input parameters used for all ML-based model generation comprise three plate dimensions (thickness, width, and length), weight density, and two elastic properties (Young's modulus Poisson's), for a total of six features. To provide a somewhat diverse—but limited—set of data used to train the models, plates were simulated using aluminum, FR-4, copper, magnesium, and stainless steel. Intrinsic and elastic properties for these materials are provided in Table 1, while the diversity of dimensions used is given in Table 2.

**Table 1.** Materials used in the generation of training data, with their intrinsic and elastic property values. Note the wide range of material properties chosen.

| Material | Weight Density (lb/in$^3$) | Young's Modulus (ksi) | Poisson's Ratio |
|---|---|---|---|
| *Aluminum* | 0.097 | 9,900 | 0.33 |
| *FR-4* | 0.070 | 2,000 | 0.12 |
| *Copper* | 0.323 | 16,000 | 0.343 |
| *Magnesium* | 0.065 | 6,500 | 0.35 |
| *Stainless Steel* | 0.286 | 29,000 | 0.27 |

**Table 2.** Dimensions of simply-supported rectangular plates used in the generation of training data. In total, five different combinations of planar dimensions were combined

with each increment of plate thickness to give a total of 500 datasets. These 500 datasets were split into 261 for training/testing and 239 for independent testing.

| Dimension | Set | Value |
|---|---|---|
| *Planar length and width, w × l* | i. | 2.000 in × 2.000 in |
| | ii. | 1.875 in × 3.000 in |
| | iii. | 5.250 in × 7.000 in |
| | iv. | 6.000 in × 3.000 in |
| | v. | 10.500 in × 8.750 in |
| *Plate Thickness, t:* | | 0.030 to 0.125 in; increments of 0.005 in |

Using the aforementioned material properties and dimensions total of 500 solutions were calculated. From this, 261 datapoints are reserved for model generation (training and testing), while the remaining 239 datapoints ae retained for independently assessing model predictive accuracy.

One final note regarding the training/testing dataset is the application of an uncertainty. To simulate a possible 2% (+/-1%) variation due to meaurement uncertainty and imperfect physical sample, the final output feature—i.e., natural frequency—is adjuted as follows:

$$f_{n,adj} = f_n(0.99 + 0.02R) \tag{1}$$

where $R$ is a random number between the values of 0 and 1 having an approximately Gaussian distribution, and $f_{n,adj}$ is the final adjusted natural frequency used at the output parameter. (For convenience, the adjusted natural frequency is simply referred to as $f_n$ for the remainder of this study.)

## 2.2. Simplified physics: natural frequency of a thin plate

Simplified physics used for integration with the deep learning model is partly based on small-deflection plate theory, adapted from Steinberg (2001). Using this approach, the fundamental natural frequency, $f_n$, of the rectangular plate simply-supported along four edges shown in Fig. 1 is as follows:

$$f_n = \frac{\pi}{2}\left(\frac{Dt}{\rho}\right)^{1/2}\left[\frac{1}{w^2} + \frac{1}{l^2}\right] \qquad (2)$$

where $t$ is the plate thickness, $\rho$ is the material mass density, $w$ and $l$ are the plate width and length, respectively, and $D$ is flexural rigidity. Flexural rigidity is computed from thickness, Young's modulus, $E$, and Poisson's ratio, $v$:

$$D = \frac{Et^3}{12(1 - v^2)} \qquad (3)$$

Note that the flexural rigidity term—which is calculated independently from planar dimensions and boundary conditions—becomes the primary physics-based relation used to augment DNN-based model generation. This is important because while the plate equation is simple enough to use to calculate fundamental frequencies for our current configuration, for the majority of real-world applications, physical boundary conditions are not easily classifiable and may be a mix of simple, clamped, or elastic, to name a few. However, the flexural modulus is relatively easy to estimate for any plate based on material elastic properties and plate thickness. Therefore, this approach can be adapted to any dimensional and boundary condition configuration.

Since flexural rigidity is only dependent on three features—plate thickness, Young's modulus, and Poisson's ratio—an attempt is made to involve the remaining dimensions and material density into simplified physics by considering the plate weight, $W$, calculated as

$$W = t \times w \times l \times \rho \tag{4}$$

Finally, shear modulus, $G$, is used as a third supplemental term:

$$G = E/[(1 + \nu)] \tag{5}$$

### 2.3. Neural network architecture

Based on early experimentation, the basic predictive model chosen and developed for this study is a deep neural network comprising four hidden layers and one output layer, as diagrammed in Fig. 3. As previously discussed, the six input features present in this basic DNN—as well as all subsequent physics-enhanced simulations—are the three plate dimensions, density, Young's modulus, and Poisson's ratio. The optimization algorithm chosen for this study is *Adam*, introduced by Kingma and Ba (2014).

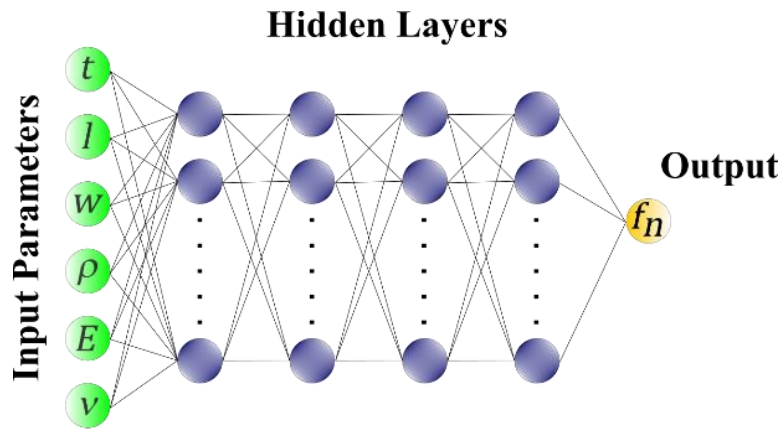

**Fig. 3.** Basic deep neural network used to predict natural frequency of thin rectangular plates simply-supported along all four edges. Note the following characteristics of this DNN: six input features (plate thickness, $t$, length, $l$, width, $w$, density, $\rho$, Young's

modulus, $E$, and Poisson's ratio, $v$); four hidden layers with several neurons each; and one output layer with a single feature, i.e., fundamental natural frequency, $f_n$.

Hyperparameter selection was decided by trial and error rather than through a rigorous optimization process. For consistency, all DNN-based model generation is conducted using the parameters provided in Table 3.

**Table 3.**     Selected hyperparameters and their values. Note that for performance comparison purposes, hyperparameters remained constant throughout the study.

| Hyperparameter | Description/Value |
| --- | --- |
| *Hidden layer activation function* | Rectified Linear Unit (ReLU) |
| *Output layer activation function* | Linear |
| *Learning rate, $\alpha_t$* | 0.01 |
| *Batch size* | 80 |
| *Epochs* | 500 |
| *Number of neurons per layer* | 40 |
| *Numerical stability parameter, $\hat{\epsilon}$* | 1e-6 |
| *Loss function* | Mean-Squared Error (MSE) |

**2.4.     Integration of physics into NN architecture**

As previously discussed, the fundamental methodology chosen for integrating physics into the process of machine learning is to insert physics-based quantities as augmentative data into various DNN layers. Four basic approaches are taken, represented in Fig. 4:

1. <u>Neural network with no physics terms.</u> This basic DNN serves as a control against which all other models are compared and contrasted. The approach taken here was introduced in Fig. 3 and is repeated in Fig. 4(a) to allow comparison with physics-based instantiations.

2. <u>Insertion of physics into Layer 1.</u> Representing the first attempt to augment the DNN, the insertion of physics first hidden layer is fundamentally equivalent to having the simplified physics-based terms integrated into the baseline dataset, as demonstrated in Fig. 4(b).

3. <u>Insertion of physics into Layers 2, 3, 4, or 5 individually.</u> With this strategy, physics is implanted directly into one of the three middle hidden layers or the output layer through concatenation with intermediate parameters generated by the preceding layer, an example of which is seen in Fig. 4(c). For the study at hand, this provides four additional scenarios.

4. <u>Insertion of physics into multiple layers for each run.</u> For this final approach to augmentation, physics terms are repeatedly inserted into the various DNN layers. For example, the same physics terms inserted into Layer 1 are added to Layers 2 through 4, and so on. Three specific cases are simulated within this framework: (i) physics in Layers 1 through 4, (ii) physics in Layers 2 through 4, shown in Fig. 4(d), and (iii) physics in Layers 2 through 5.

These four strategies represent nine total DNNs: one basic network without additional physics terms and eight different ways of integrating physics into the baseline DNN.

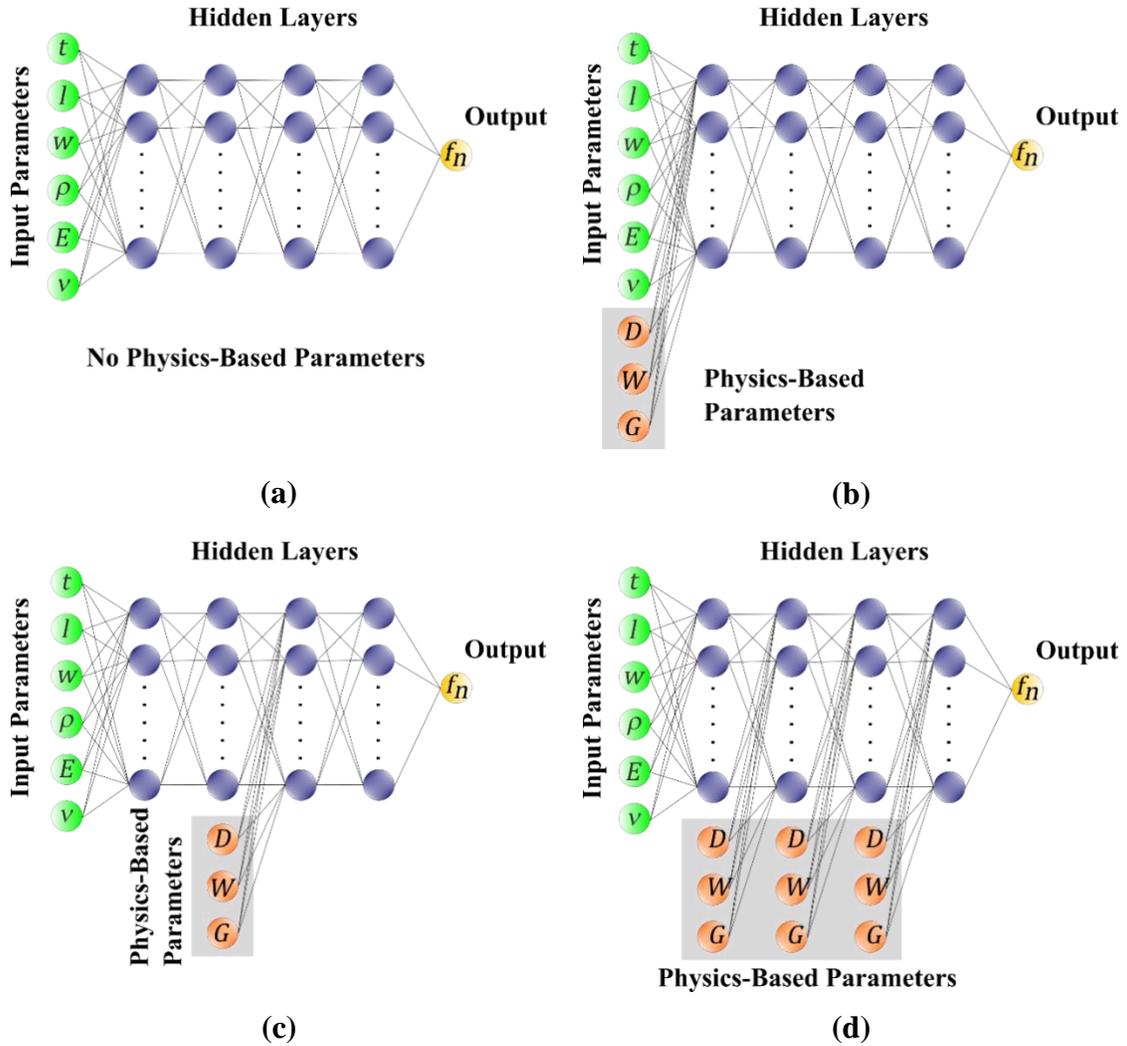

**Fig. 4.** Examples of deep neural networks used to generate predictive models for fundamental resonance of a rectangular plate simply-supported along all edges. (a) DNN driven purely by data. This network is identical to that shown in Fig. 3. (b) DNN with augmented data set generated from physical relationships among original data. Here the physics-based parameters are simply added to the baseline inputs to give up to nine total input features. (c) DNN with physics integrated into a single layer (Layer 3 shown). In this architecture, physics-based features are added to a hidden layer as additional inputs for that layer only. (d) DNN with physics integrated into multiple layers (Layers 2 through 4 shown). With this final example, physics-based features are inserted into an early hidden layer and re-inserted at one or more subsequent layers.

Along with varying locations of physics integration within the DNN model, six different combinations of calculated parameters are examined, listed in Table 4. The purpose behind this is to observe whether single or multiple variables show the most improvement when integrated into the DNN.

**Table 4.** Various models developed to predict natural frequency. Specifically, this entails the injection of different combinations of weight ($W$), flexural modulus ($D$), and shear modulus ($G$) at each DNN layer.

| Model | Physics | Code File Name |
|---|---|---|
| *Baseline Deep Neural Network* | None | NNML.py |
| *Physics-Guided Deep Neural Network 1* | $W, D$ | PGML.py |
| *Physics-Guided Deep Neural Network 2* | $W$ | PGMLa.py |
| *Physics-Guided Deep Neural Network 3* | $D$ | PGMLb.py |
| *Physics-Guided Deep Neural Network 4* | $G$ | PGMLc.py |
| *Physics-Guided Deep Neural Network 5* | $W, D, G$ | PGMLd.py |
| *Physics-Guided Deep Neural Network 6* | $D, G$ | PGMLf.py |

The final consideration for model development is amount of training data available, for which the following dataset sizes are used: 261 datapoints, 117 datapoints, 60 datapoints, and 30 datapoints. The largest dataset containing 261 points serves as a superset from which the 117, 60, and 30 are derived. The goal in examining different training dataset sizes is to determine how much—if any—impact physics has on improving accuracy of the DNN when only scarce data are available.

Based on the various combinations described herein, the total number of models generated for this study is

$$(9 \text{ insertion options}) \times (7 \text{ physics combinations}) \\ \times (4 \text{ training data sizes}) = \mathbf{252\ models} \tag{6}$$

Each of the final 252 models is used to compute multiple predictions of fundamental natural frequency while initialized with different random seeds, where the final output value is realized as an average of all computed values (each having equal weight). The purpose for this is to reduce variation in model predictions. In total, fifty values of natural frequency are computed for each model. This number of averages was chosen based on early trials demonstrating that the overall average would tend toward a quasi-stable value. An example of this trend is shown in Fig. 5

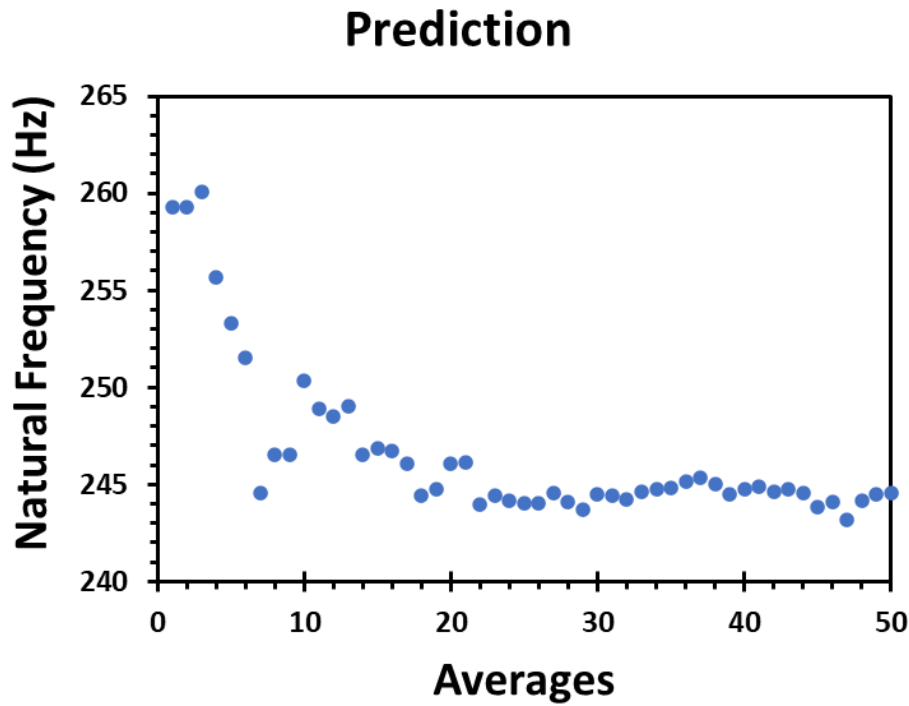

**Fig. 5.** Example of averaging used for final model predictions. Note that as the number of averages increases, the final natural frequency value tends to stabilize at about 245 Hz. This approach is used to reduce the amount of variation within natural frequency predictions. (Results shown are for baseline DNN predictions of a stainless-steel plate with dimensions 5.25 in x 7.00 in x 0.045 in.)

## 2.5. Error

The final term of high importance computed for assessment of model accuracy is percentage error. Once a solution of fifty averages is attained, the error for each resultant datapoint is calculated as

$$Error = \frac{|f_{n,a} - f_{n,p}|}{f_{n,a}} \times 100\% \tag{7}$$

where $f_{n,a}$ is the actual natural frequency of the plate and $f_{n,p}$ is the natural frequency predicted by the DNN-based model. Based on this information, several basic quantities are available: mean error, median error, standard deviation of error, and maximum and minimum error. However, for comparative consistency and simplicity, mean error is chosen as the metric used in accuracy comparisons.

### 2.6. Independent assessment of physics-guided DNN model accuracy

Following development of predictive models through training and testing with original data having parameters from Table 1 and Table 2, each is tested using two independent datasets in order to assess prediction accuracy. The first dataset independently tested with each model—henceforth referred to as *Test Dataset 1*—contains the remaining 239 datapoints developed from finite element modeling of the aluminum, copper, FR-4, magnesium, and stainless steel plates. The second independent dataset—*Test Dataset 2*—consists of a series of 101 plates constructed from composite printed wiring board (PWB) material having the isotropic property values and random dimensions within the extreme parameters provided in Table 5.

**Table 5.** Independent Dataset 2 input features. Material elastic properties taken from Steinberg (2001), while density is estimated. Planar length (*l*), planar width (*w*), and plate thickness (*t*) for the 101 different plates are generated at random within the bounds shown.

| Feature | Value |
| --- | --- |
| *Density* | 0.150 lb/in$^3$ |
| *Young's modulus* | 3,000 ksi |
| *Poisson's ratio* | 0.18 |
| *Planar length, l* | 2.040 to 9.824 in |
| *Planar width, w* | 2.016 to 9.843 in |
| *Plate Thickness, t:* | 0.024 to 0.216 in |

## 3. Results and discussion

### 3.1. Natural frequency predictions using data similar to training

Fig. 6 shows mean predictive error for each physics-guided DNN model when compared with the baseline DNN containing no physics. Test Dataset 1 is used for independent testing of each model. Note that mean error for the baseline DNN without physics is 2.1% for the largest training size of 261 datapoints, which is approximately equivalent to the range of uncertainty assumed for plate natural frequency measurements. As expected, baseline DNN models generated with decreased training dataset sizes of 117, 60, and 30 datapoints exhibit increasingly higher mean error, with values of 5.0%, 6.4%, and 19.3%, respectively.

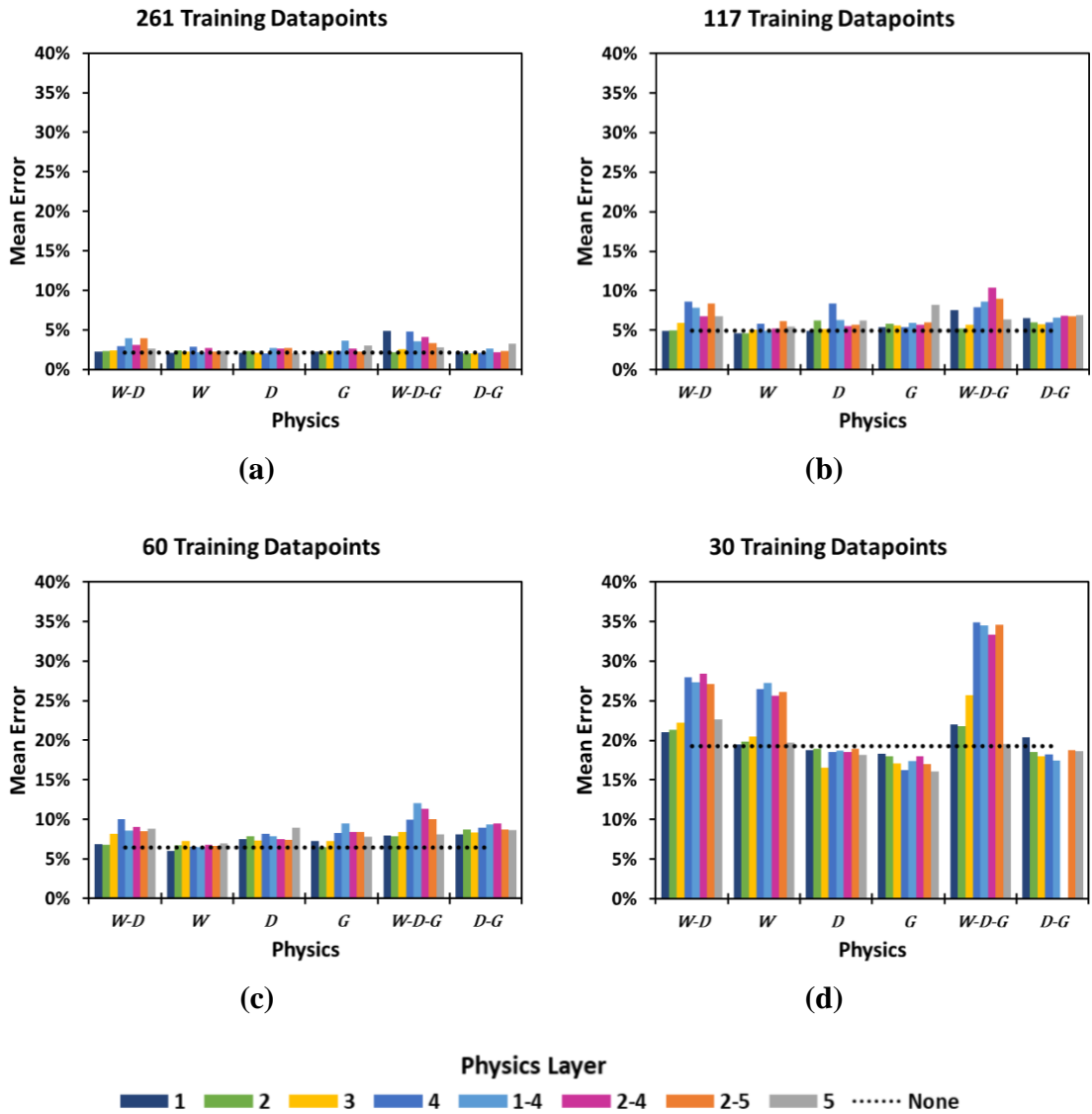

**Fig. 6.** Mean prediction error for Training Dataset 1 using models created by physics-enhanced DNNs generated with (a) 261 training samples, (b) 117 training samples, (c) 60 training samples, and (d) 30 training samples. For each layer of augmentation, several different combinations of plate weight (*W*), flexural rigidity (*D*), and material shear modulus (*G*) are evaluated.

Two items immediately stand out from the data. The first observation is that insertion of physics into the DNNs results in virtually no improvement for models generated with 261, 117, and 60 datapoints. In fact, for these simulations error slightly increases above the baseline DNN in most cases and almost doubles for the weight-flexural modulus-shear modulus combination with 117 training datapoints. The second observation, however, is that minor improvements are seen for models generated with 30 datapoints when flexural rigidity, shear modulus, or the combination of the two are used. In spite of this minimal improvement for the scarcest dataset, in all four cases several physics-enhanced models actually produce results worse than the baseline. In particular, mean error is observed to increase by approximately 80% when the combination of plate weight, flexural modulus, and shear modulus are used as augmentative features for the model trained with thirty datapoints. Overall, no notable or consistent benefit is evident from the inclusion of physics-based parameters when testing with data similar to that used for training.

### 3.2. Natural frequency predictions using data dissimilar from training

**Error! Reference source not found.** compares mean predictive error for the physics-guided DNN models to the baseline DNN when independently tested with Test Dataset 2.

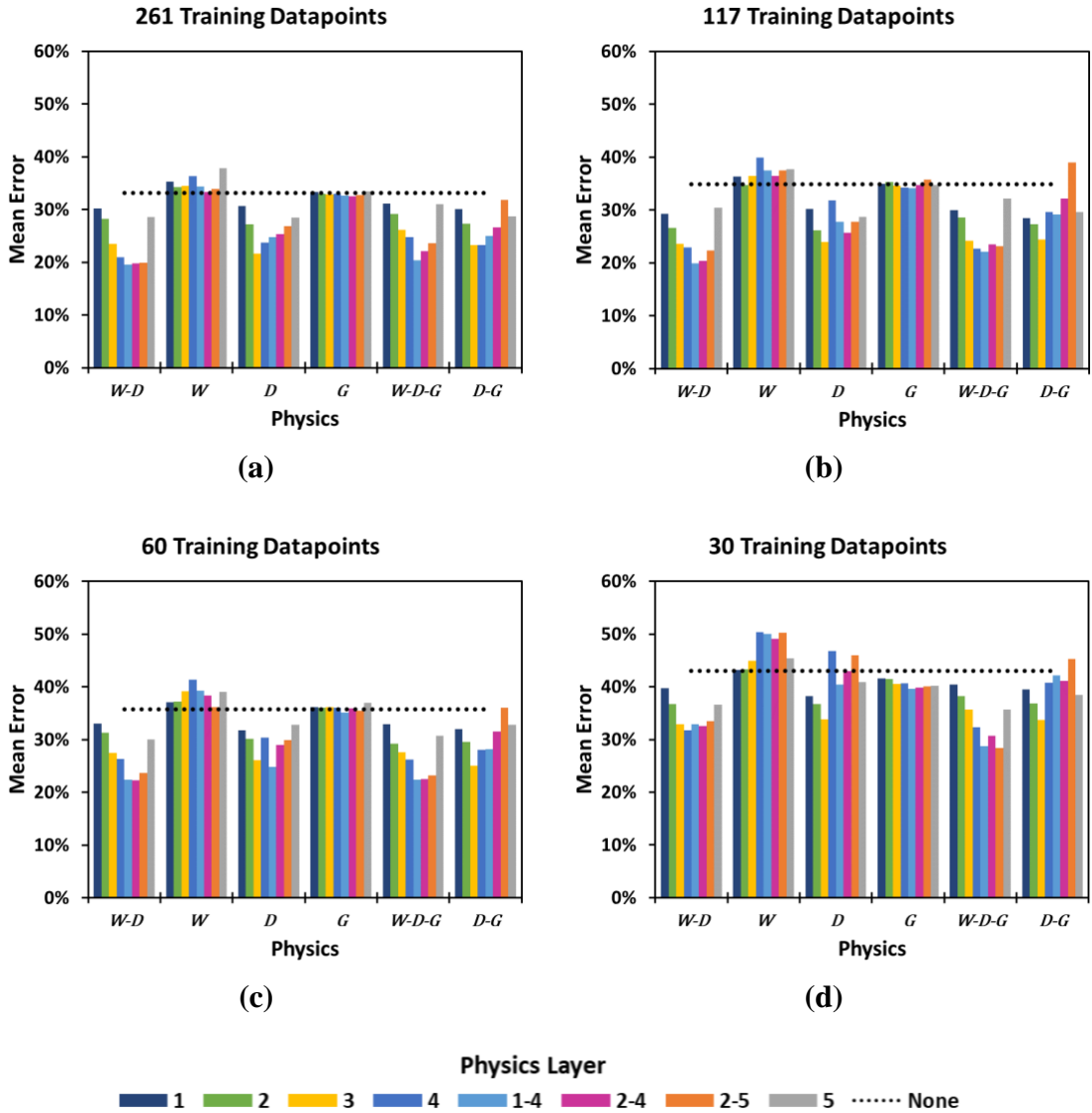

**Fig. 7.** Mean prediction error for Training Dataset 2 using models created by physics-enhanced DNNs generated with (a) 261 training samples, (b) 117 training samples, (c) 60 training samples, and (d) 30 training samples. For each layer of augmentation, several different combinations of plate weight (*W*), flexural rigidity (*D*), and material shear modulus (*G*) are evaluated.

In contrast to independent Test Dataset 1, although less accurate overall, Test Dataset 2 sees moderate but clear improvement to natural frequency predictions for certain combinations of physics parameters and insertion layers covering all four training dataset sizes. In particular, solutions where flexural rigidity is included as a physical parameter almost universally demonstrate better accuracy of natural frequency prediction over the baseline DNN without physics. For example, the greatest increase in accuracy occurs with inclusion of weight and flexural modulus in DNN Layers 1 through 4, exhibiting a total predictive error reduction of 15.0% (from 33.2% to 19.6%) when the training dataset contains 117 points.

Interestingly, while not as effective when inserted into the DNN with larger training datasets, the addition of shear modulus to the W-D combination exhibits similar performance to just weight and flexural modulus alone for the training dataset of size 60, and the W-D-G combination actually outperforms all others when the model is trained with only thirty 30 datapoints. Although not conclusive, this may indicate that certain combinations of physics-based parameters are more effective than others when applied to different training dataset sizes during model generation.

### 3.3. Loss and overfitting

One aspect of this study not yet discussed in detail is the comparison of training and validation loss. During the course of the investigation, it is observed that as the training dataset sizes decrease, the training and validation loss (based on mean-squared error, as indicated in Table 3) these two tracking quantities diverge from one another, as is evident in the example shown in Fig. 8. Although expected behavior for the DNN where all hyperparameters are held constant, it likely indicates an increasing degree of overfitting. As such, the number of epochs used for this study is too large and reduction to something less than 100 would improve computational efficiency while not likely deteriorating accuracy.

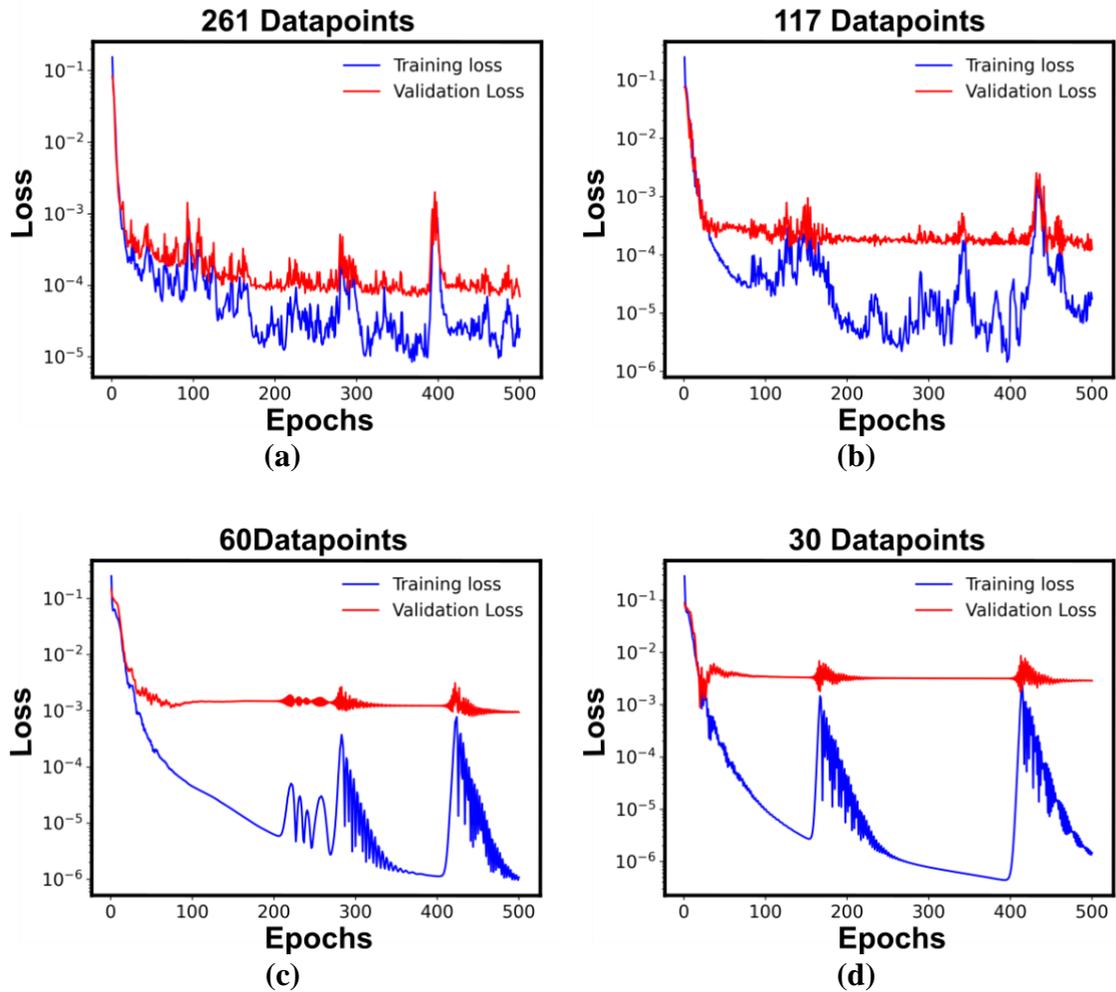

**Fig. 8.** Examples of training and validation loss for (a) 261 training samples, (b) 117 training samples, (c) 60 training samples, and (d) 30 training samples. Note that while the loss representing the highest training size of 261 points is fairly well-behaved, as the training size is decreased, the validation loss quickly diverges from the training loss after very few epochs. This behavior coincides with the loss in predictive accuracy for models trained with scarcer data, and it likely indicates an overfitting condition (Loss plots taken from model generated with flexural modulus inserted into layers 2, 3, and 4.)

### 3.4. General observations

## 4. Conclusions

While the physical phenomenon chosen is resonance of thin rectangular plates simply-supported on all edges, this study serves as a continuation of an earlier investigation that improved behavioral predictions of airfoils through introduction of simplified physics into a single internal neural network layer (Pawar et al., 2021). To build on the momentum of that research, physics-enhanced parameters informing the deep neural network described herein are initially injected into each DNN layer one-at-a-time, first as supplemental input features (Layer 1), then sequentially into the three internal layers (Layers 2, 3, and 4), and finally at the output (Layer 5). In addition to these five case studies, three further scenarios are evaluated where the same physics-enhanced features are reinserted into multiple layers during a single DNN computation. When tested by means of independent data, supplementation with simplified physics-based parameters provides virtually no reduction in prediction error over the baseline for models trained with dataset sizes of 60 and greater, although a small improvement of slightly better than 3% is observed when trained with a sparser size of 30 and physics introduced either to Layer 4 or 5 (but not multiple layers). However, notable gains in accuracy occur when the independent test data is for material and dimensions not conforming to the training set. Particularly, reductions in error from 33.2% to 19.6%, 34.9% to 19.9%, 35.8% to 22.4%, and 43.0% to 28.4% are achieved for training dataset sizes of 261, 117, 60, and 30, respectively. For these cases, injection of physics into multiple layers per DNN consistently outperforms instances where only one layer is augmented, and this disparity is more apparent for models trained with 30 points. The initial lack of error reduction for similar training/independent test datasets coupled with subsequent greater improvements for dissimilar independent test data suggests that the approach described herein indeed provides improvement to DNN-based model generalizability. Future evaluation of this methodology may show that hyperparameter optimization further improves predictive accuracy, or that a combination of hyperparameter tuning and loss function tailoring is ideal.

**Acknowledgments**

This work was partially supported by the Office of Naval Research Grant N000141712620 and National Science Foundation Award 2014346.

**Conflict of Interest**

The authors declare no conflict of interest.